\documentclass[aps,pra,reprint,groupedaddress]{revtex4-1}

\usepackage{bbm, bm, epsfig, amsmath, amsfonts, subfigure, amssymb, wasysym, graphicx}
\usepackage[hypertex, bookmarks=false, pdfstartview=FitH]{hyperref}

\begin{document}


\title{Thermodynamics of balanced and slightly spin-imbalanced Fermi gases at unitarity}



\author{Olga Goulko}%
\author{Matthew Wingate}%
\affiliation{%
Department of Applied Mathematics and Theoretical Physics, University of Cambridge, Centre for Mathematical Sciences, Cambridge CB3 0WA, United Kingdom}%



\date{\today}

\begin{abstract}
In this paper we present a Monte Carlo calculation of the critical temperature and other thermodynamic quantities for the unitary Fermi gas with a population imbalance (unequal number of fermions in the two spin components). We describe an improved worm type algorithm that is less prone to autocorrelations than the previously available methods and show how this algorithm can be applied to simulate the unitary Fermi gas in presence of a small imbalance. Our data indicates that the critical temperature remains almost constant for small imbalances $h=\Delta\mu/\varepsilon_F\lessapprox0.2$. We obtain the continuum result $T_c=0.171(5)\varepsilon_F$ in units of Fermi energy and derive a lower bound on the deviation of the critical temperature from the balanced limit, $T_c(h)-T_c(0)>-0.5\varepsilon_Fh^2$. Using an additional assumption a tighter lower bound can be obtained. We also calculate the energy per particle and the chemical potential in the balanced and imbalanced cases.
\end{abstract}


\maketitle

\relpenalty=9999
\binoppenalty=9999

\renewcommand{\Im}{\textnormal{Im}}
\newcommand{\sgn}{\textnormal{sign}}
\newcommand{\Tr}{\textnormal{Tr}}

\section{Introduction}
The Fermi gas at unitarity -- a dilute system of two-component fermions interacting with divergent scattering length -- is a particularly interesting example of a strongly interacting fermionic system \cite{review1, review2, rev_koehler}. In this case the density sets the only relevant length scale and the system exhibits universal behaviour. At a certain critical temperature a phase transition into a superfluid state takes place. In contrast to the weakly interacting BCS limit the critical temperature at unitarity is of order of the Fermi temperature, and hence accessible for experimental study. Examining the mechanism behind this phase transition promises valuable insights into high-temperature superfluidity.

In the strongly interacting limit, perturbative methods are inapplicable and mean-field approaches involve uncontrolled approximations, since the contribution from fluctuations becomes significant \cite{review2}. Hence, numerical methods have found wide application. The method underlying our study is the Diagrammatic Determinant Monte Carlo (DDMC) algorithm \cite{burovski}, which was developed for the calculation of the critical temperature and was specifically designed to take advantage of the physical properties in the unitarity limit. Here we will describe our tests and implementation of this algorithm and suggest several modifications that significantly reduce autocorrelations and therefore increase the efficiency. Preliminary results were presented in \cite{goulko}.

So far, most numerical studies were limited to the balanced case, when the number of fermions in the two spin components is equal. An imbalance in the particle number results in new interesting effects, which makes a detailed numerical study desirable \cite{reviewimbalanced, reviewimbalanced2}. On the other hand, an imbalance also leads to algorithmic difficulties, which are due to fermionic statistics. The Feynman diagrams in the expansion of the partition function have different signs, depending on the number of fermionic loops. In the balanced case this problem can be avoided, since the diagrams of each order can be represented as a square of a matrix determinant \cite{rubtsov}. In the presence of an imbalance this is no longer the case, so that the partition function cannot be used as a probability distribution for Monte Carlo sampling. A similar problem arises in lattice QCD, where a non-zero chemical potential renders the fermionic determinant complex. Many techniques dealing with the sign problem have been developed, the most straightforward of which is the ``sign quenched method'' \cite{phasequenched}, which we will make use of in our study. We found that at unitarity the sign problem is mild, which is also consistent with the observation that the superfluid state remains remarkably stable in response to increasing imbalance.

In this work we present a numerical calculation of the critical temperature and other thermodynamic observables for several values of imbalance. We begin with a review of the Fermi-Hubbard model and the finite temperature formalism in Sec.~\ref{modelsection}. In Sec.~\ref{ordersection} we define the order parameter for the phase transition and show how the critical temperature can be extracted from the numerical data. We then summarise our version of the worm algorithm in Sec.~\ref{wormsection} and explain how it can be generalised to the imbalanced case. Finally, our results for the balanced as well as the imbalanced gases are presented and discussed in Sec.~\ref{resultssection}.

\section{Fermi-Hubbard model at finite temperature\label{modelsection}}
The Fermi-Hubbard model is the simplest lattice model for two-particle scattering. Its Hamiltonian in the grand canonical ensemble is given by
\begin{equation}
H=H_0+H_1=\sum_{\mathbf{k},\sigma}(\epsilon_\mathbf{k}-\mu_\sigma)c^\dagger_{\mathbf{k}\sigma}c_{\mathbf{k}\sigma}+U\sum_{\mathbf{x}}c^\dagger_{\mathbf{x}\uparrow}c_{\mathbf{x}\uparrow}c^\dagger_{\mathbf{x}\downarrow}c_{\mathbf{x}\downarrow},
\end{equation}
where $\epsilon_\mathbf{k}=\frac{1}{m}\sum_{j=1}^{3}(1-\cos{k_j})$ is the discrete dispersion relation, and $c^\dagger_{\mathbf{k}\sigma}$ ($c_{\mathbf{k}\sigma}$) the time-dependent fermionic creation (annihilation) operator. We set $\hbar=k_B=1$ throughout. We have chosen this simple dispersion relation for a better comparison with reference \cite{burovski} where the same relation was used. It is possible to speed the approach to the continuum limit by choosing a more complex dispersion relation \cite{disprels}. This will be explored in future work. This model describes non-relativistic fermions of two species labelled by $\sigma$ (which we will call ``spin up'' and ``spin down'') with equal particle mass $m$. The attractive contact interaction is characterised by the coupling constant $U<0$. This coupling can be tuned so that the scattering length takes infinite value, by solving the two-body problem in the same way as it was done in \cite{burovski}. The corresponding value is $U=-7.914$, in units where $m=1/2$. We work on a 3D simple cubic spatial lattice with $L^3$ sites, periodic boundary conditions and lattice spacing set to unity. The continuum limit of this model can be taken by extrapolation to vanishing filling factor $\nu=\langle \sum_{\sigma}c^\dagger_{\mathbf{x}\sigma}c_{\mathbf{x}\sigma}\rangle\rightarrow0$.

To study the finite temperature behaviour we work with the grand canonical partition function in the imaginary time interaction picture, $Z=\Tr e^{-\beta H}$, where $\beta$ is the inverse temperature. The imaginary time direction remains continuous. Using Dyson's formula and expanding $Z$ in powers of $H_1$ generates a series of Feynman diagrams, where each 4-point vertex has one incoming line of each spin and one outgoing line of each spin. The Feynman rules assign a factor of $(-U)$ to a vertex and a line represents a free (finite temperature) single-particle propagator,
\begin{eqnarray}
G_{(0)}^\sigma(\mathbf{x}_i-\mathbf{x}_j,\tau_i-\tau_j)\!\equiv&&-\langle\mathbf{T}_\tau c^\dagger_{\mathbf{x}_i\sigma}(\tau_i)c_{\mathbf{x}_j\sigma}(\tau_j)\rangle\qquad\qquad\quad\\
=&&-\Tr[\mathbf{T}_\tau e^{-\beta H_0}c^\dagger_{\mathbf{x}_i\sigma}(\tau_i)c_{\mathbf{x}_j\sigma}(\tau_j)],
\label{propagatordef}
\end{eqnarray}
where $\mathbf{T}_\tau$ denotes the imaginary time ordering operator. The explicit form of the propagator in momentum space is given by \cite{LL10}
\begin{equation}
G^{\sigma}_{(0)}(\mathbf{k},\tau\equiv\tau_j-\tau_i)=
\left\{\begin{array}{lll}
	e^{-(\epsilon_{\mathbf{k}}-\mu_\sigma)\tau}(1-n_{\mathbf{k}\sigma}) 	& \mbox{for} & \tau>0 \\
	-e^{-(\epsilon_{\mathbf{k}}-\mu_\sigma)\tau}n_{\mathbf{k}\sigma} 	& \mbox{for} & \tau\leq0
\end{array}\right.\!\!,
\label{propagatorexplicit}
\end{equation}
where $n_{\mathbf{k}\sigma}=(1+e^{\beta(\epsilon_{\mathbf{k}}-\mu_\sigma)})^{-1}$ is the occupation of the state $(\mathbf{k},\sigma)$ for free fermions. Additionally, each fermionic loop contributes a minus sign, with the consequence that the diagrams in the series have different signs. Since we are ultimately interested in thermal expectation values of operators and thermal averages are calculated using the expansion of the partition function, it would be convenient to use this expansion as a probability distribution to generate configurations for Monte Carlo sampling. For this purpose we need to rewrite the series as a sum of positive terms only. It was shown in \cite{rubtsov} that the partition function can be written as
\begin{equation}
Z=\sum_{S_p}(-U)^p\det\mathbf{A}^{\uparrow}(S_p)\det\mathbf{A}^{\downarrow}(S_p),
\label{partitionfunction}
\end{equation}
where $S_p$ denotes a vertex configuration (the spacetime positions of all vertices) and the matrix entries are the propagators $A^\sigma_{ij}(S_p)=G_{(0)}^\sigma(\mathbf{x}_i-\mathbf{x}_j,\tau_i-\tau_j)$, given by Eqs.~(\ref{propagatordef}) and (\ref{propagatorexplicit}). Note that within this formalism the only degrees of freedom are the vertex coordinates and hence it is not necessary to distinguish between the different ways of connecting them. If the chemical potential is equal for spin up and spin down fermions (the balanced case) we have $\det\mathbf{A}^\uparrow\det\mathbf{A}^\downarrow=|\det\mathbf{A}|^2$, so that all terms in the series are positive.

\section{Order parameter and finite-size scaling\label{ordersection}}
\begin{figure}
\includegraphics[width=\columnwidth]{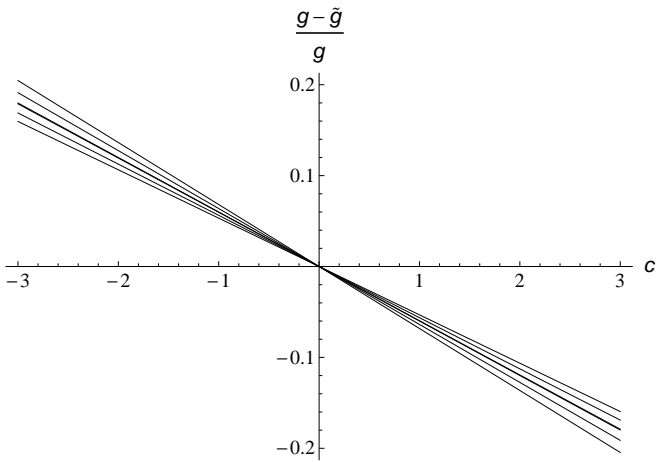}
\caption{\label{scalingcomp}The relative difference $(g(L_i,L_j)-\tilde{g}(L_i,L_j))/g(L_i,L_j)$ as a function of $c$ (in lattice units), for several values of $L_i$ and $L_j$ ranging between $10$ and $16$.}
\end{figure}

The physical observable in the focus of our study is an order parameter for the phase transition to superfluidity. To define the order parameter, which is related to the density of the condensate, we first introduce the pair creation and annihilation operators $P^\dagger(\mathbf{x}',\tau')=c^\dagger_{\mathbf{x}'\uparrow}(\tau')c^\dagger_{\mathbf{x}'\downarrow}(\tau')$ and $P(\mathbf{x},\tau)=c_{\mathbf{x}\uparrow}(\tau)c_{\mathbf{x}\downarrow}(\tau)$. At the critical point the correlation function
\begin{eqnarray}
G_2(\mathbf{x}\tau;\mathbf{x}'\tau')&=&\left\langle\mathbf{T}_\tau P(\mathbf{x},\tau)P^\dagger(\mathbf{x}',\tau')\right\rangle\\
&=&\frac{1}{Z}\textnormal{Tr}[\mathbf{T}_\tau P(\mathbf{x},\tau)P^\dagger(\mathbf{x}',\tau')e^{-\beta H}]
\end{eqnarray}
is proportional to $|\mathbf{x}-\mathbf{x}'|^{-(1+\eta)}$ as $|\mathbf{x}-\mathbf{x}'|\rightarrow\infty$ (in three dimensions), where $\eta\approx0.038$ \cite{critexp, critexpreview} is the anomalous dimension for the U$(1)$ universality class. Hence, if no corrections due to irrelevant operators were present, the rescaled integrated correlation function
\begin{equation}
R(L,T)=L^{1+\eta}(\beta L^3)^{-2}\sum_{\mathbf{x},\mathbf{x}'}\int_{0}^{\beta}d\tau\int_{0}^{\beta}d\tau'G_2(\mathbf{x}\tau;\mathbf{x}'\tau')
\label{intcorrfn}
\end{equation}
would be independent of lattice size $L$ at the critical $\beta_c=1/T_c$ \cite{burovski}. In this case, all $R(L,T)$ curves for different values of $L$ would cross in a single point. Taking scaling violations into account, the function $R(L,T)$ near the critical point can be written as a product of a universal analytic scaling function $f(x)$, where $x=L^{1/\nu_\xi}t$, and a correction term due to finite lattice size,
\begin{equation}
R(L,T)=f(L^{1/\nu_\xi}t)(1+cL^{-\omega}+\ldots).
\label{Rscaling}
\end{equation}
Here $t=(T-T_c)/T_c$, $c$ is a non-universal constant, and the critical exponents $\omega\approx0.8$ and $\nu_\xi\approx0.67$ can be determined to high precision with various methods, see e.g. \cite{critexp, critexpreview}. Near the critical point we can expand Eq.~(\ref{Rscaling}) and keeping only terms linear in $t$ we obtain
\begin{equation}
R(L,T)=(f_0+f_1(T-T_c)L^{1/\nu_\xi}+\ldots)(1+cL^{-\omega}+\ldots).
\label{Rscalingexp}
\end{equation}
Previous work \cite{burovski,bulgac1} used a two-step procedure for determining $T_c$ from this equation. First the crossings of the $R(L,T)$ curves for each pair of lattice sizes were determined individually. By equating $R(L_i,T_{ij})=R(L_j,T_{ij})$ and using Eq.~(\ref{Rscalingexp}), a relation between $T_c$ and the crossing temperatures $T_{ij}$ can be derived,
\begin{equation}
T_{ij}-T_c=\kappa g(L_i,L_j),
\end{equation}
where
\begin{equation}
g(L_i,L_j)=\frac{(L_j/L_i)^\omega-1}{L_j^{\frac{1}{\nu_\xi}+\omega}\left(1-(\frac{L_i}{L_j})^{\frac{1}{\nu_\xi}}\right)+cL_j^{\frac{1}{\nu_\xi}}\left(1-(\frac{L_i}{L_j})^{\frac{1}{\nu_\xi}-\omega}\right)},
\label{Rscalinglin}
\end{equation}
and $\kappa=cf_0/f_1$ is a non-universal constant. In the references \cite{burovski} and \cite{bulgac1} the second term in the denominator of $g(L_i,L_j)$ is neglected, so that the second step of the procedure simplifies to a linear fit of the $T_{ij}$ to the function
\begin{equation}
\tilde{g}(L_i,L_j)=\frac{(L_j/L_i)^\omega-1}{L_j^{\frac{1}{\nu_\xi}+\omega}\left(1-(\frac{L_i}{L_j})^{\frac{1}{\nu_\xi}}\right)}.
\end{equation}
The critical temperature is the intercept of this fit. This simplification is justified if the constant $c$ is sufficiently small. But already if $c$ assumes values of order of unity the systematic error associated with this approximation can reach up to $20\%$, as shown in Fig.~\ref{scalingcomp}.
\begin{figure}
\includegraphics[width=\columnwidth]{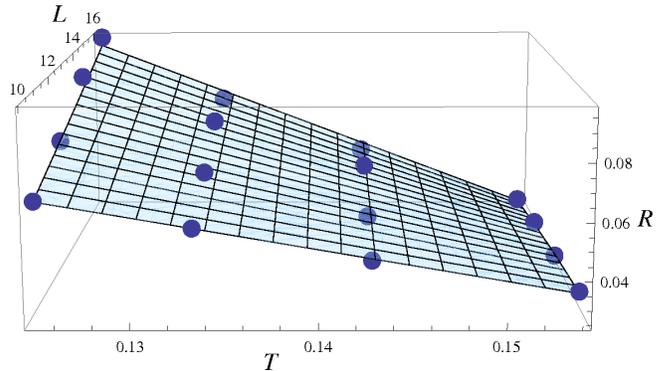}
\caption{\label{fit56dataan}(Color online) A typical fit of the rescaled correlation function $R(L,T)$ according to Eq.~(\ref{Rscalingexp}). Here data was taken at four different lattices sizes and temperatures. For this fit $\chi^2/$d.o.f$=1.4$. The value for $c$ was found to be $-1.4(5)$. All quantities are given in lattice units.}
\end{figure}
To avoid this systematic uncertainty we choose a different procedure for extracting $T_c$ from the numerical data. In our analysis we use Eq.~(\ref{Rscalingexp}) directly to fit all data triplets $(R,L,T)$ to a single function. An example of such a non-linear fit is shown in Fig.~\ref{fit56dataan}. This procedure has several advantages. Firstly, the previously described systematic error is no longer present, which can become relevant, since we found $|c|>1$ in several cases (see Sec.~\ref{resultssection}). Secondly, all information obtained from the simulations is used for the data analysis. In the original two-step procedure the $(R,T)$ tuples for each $L$ were fitted to a line separately, which involved two unknown parameters for each value of $L$. After the crossings of these lines were determined, the information about the values of $R$ could no longer be used for the next stage of the analysis. From the crossings another linear fit involving two unknown parameters had to be made. For our example from Fig.~\ref{fit56dataan} with $16$ datapoints, the original procedure would require four independent linear fits ($8$ parameters) and another linear fit into which the errors of the previous fits propagate. The new method suggested here only requires a single non-linear fit of $4$ parameters: $f_0$, $f_1$, $c$ and $T_c$, of which only $T_c$ is of interest here.

The value for $T_c$ is obtained in lattice units and needs to be translated into physical units. Since the only physical length scale at unitarity is determined by the density, the corresponding physical quantity has to be $T_c/\varepsilon_F$, where the Fermi energy is defined as $\varepsilon_F=(3\pi^2\nu)^{2/3}$. In the grand canonical ensemble the chemical potential is fixed and the corresponding filling factor $\nu$ is measured for different values of lattice size. For sufficiently large lattices the values $\nu(L)$ scale linearly with $1/L$ and an extrapolation to $1/L\rightarrow0$ will yield the thermodynamic limit for the filling factor at a given chemical potential. Finally, the continuum limit for the critical temperature is taken by extrapolating to $\nu\rightarrow0$ \cite{burovski, chenkaplan}.

\section{Implementing the algorithm\label{wormsection}}
\subsection{Balanced case}
The configuration space of diagrams can be sampled via a Monte Carlo Markov chain process: in each step one of the possible updates to another vertex configuration is proposed with probability $W(S_p\rightarrow S'_q)$ and accepted with probability $P(S_p\rightarrow S'_q)=\textnormal{min}(1,\mathcal{R})$, given by the detailed balance equation
\begin{equation}
\mathcal{R}W(S_p\rightarrow S'_q)\mathcal{D}^{(Z)}(S_p)=W(S_p\leftarrow S'_q)\mathcal{D}^{(Z)}(S'_q),
\end{equation}
where $\mathcal{D}^{(Z)}(S_p)=(-U)^p|\det\mathbf{A}(S_p)|^2$ stands for the diagram corresponding to the vertex configuration $S_p$. The requirements of detailed balance and ergodicity ensure that the configurations produced are indeed distributed according to the correct thermal probability distribution $\rho_Z(S_p)=\frac{1}{Z}(-U)^p|\det\mathbf{A}(S_p)|^2$.

The Monte Carlo estimator for a generic thermodynamic observable $\langle\hat{X}\rangle=\frac{1}{Z}\Tr[\hat{X}e^{-\beta H}]$ can also be found easily. If we denote the diagrams in the expansion of $\Tr[\hat{X}e^{-\beta H}]$ by $\mathcal{D}^{(X)}(S_p)$, as we did for the diagrams in the expansion of the partition function, we can write
\begin{eqnarray}
\langle\hat{X}\rangle=\frac{1}{Z}\sum_{S_p}\mathcal{D}^{(X)}(S_p)&=&\sum_{S_p}\frac{\mathcal{D}^{(X)}(S_p)}{\mathcal{D}^{(Z)}(S_p)}\rho_Z(S_p)\\
&\equiv&\langle\mathcal{Q}^{(X,Z)}(S_p)\rangle_{\rho_Z}.
\label{MCestbal}
\end{eqnarray}
Here $\mathcal{Q}^{(X,Z)}(S_p)\equiv\frac{\mathcal{D}^{(X)}(S_p)}{\mathcal{D}^{(Z)}(S_p)}$ is the desired Monte Carlo estimator, given by the ratio of the weights, and $\langle\ldots\rangle_{\rho_Z}$ stands for averaging over a sequence of Monte Carlo vertex configurations $S_p$ created according to the probability distribution $\rho_Z(S_p)$.

The diagrammatic expansion of the correlation function $\Tr[\mathbf{T}_\tau P(\mathbf{x},\tau)P^\dagger(\mathbf{x}',\tau')e^{-\beta H}]$ is similar to that of the partition function $Z$, but contains an additional pair of 2-point vertices at $(\mathbf{x},\tau)$ and $(\mathbf{x}',\tau')$. It is thus of advantage to sample these two series in the same simulation. In addition to sampling the regular 4-point diagrams we allow updates that insert the pair of 2-point vertices (``worm vertices'') into the configuration space \cite{burovski}. In matrix notation this means that if the worm vertices are present, each of the Green's function matrices $\mathbf{A}^\sigma$ gets an additional row and column, coming from contractions with $P^\dagger$ and $P$ respectively. We denote the extended probability distribution by $\rho_W(S_p)=\frac{1}{Z_W}(-U)^p|\det\mathbf{A}(S_p)|^2$, where the space of possible vertex configurations $S_p$ has been enlarged. This yields a change in the normalisation constant, such that the (extended) partition function now takes the form
\begin{equation}
Z_W=Z\left(1+\zeta\sum_{\mathbf{x},\mathbf{x}'}\int_{0}^{\beta}d\tau\int_{0}^{\beta}d\tau'G_2(\mathbf{x}\tau;\mathbf{x}'\tau')\right),
\end{equation}
where $\zeta$ is an arbitrary parameter. The advantage is that the Monte Carlo estimator for the order parameter now becomes very simple: it is just a constant times the ratio of configurations with and without worm vertices. Other physical observables like the number density or the energy are still only measured when the system is in the ``physical sector'', namely when the worm vertices are not present. Due to the extension of the domain and the corresponding change in the normalisation constant we need to introduce an additional rescaling:
\begin{equation}
\langle\hat{X}\rangle=\left\langle\mathcal{Q}^{(X,Z)}\right\rangle_{\rho_Z}=\frac{Z_W}{Z}\left\langle\mathcal{Q}^{(X,Z_W)}\right\rangle_{\rho_W}=\frac{\left\langle\mathcal{Q}^{(X,Z_W)}\right\rangle_{\rho_W}}{\left\langle\mathcal{Q}^{(Z,Z_W)}\right\rangle_{\rho_W}}.
\label{MCestbalext}
\end{equation}

\begin{figure*}
\hspace{0.02\textwidth}
\includegraphics[width=.46\textwidth]{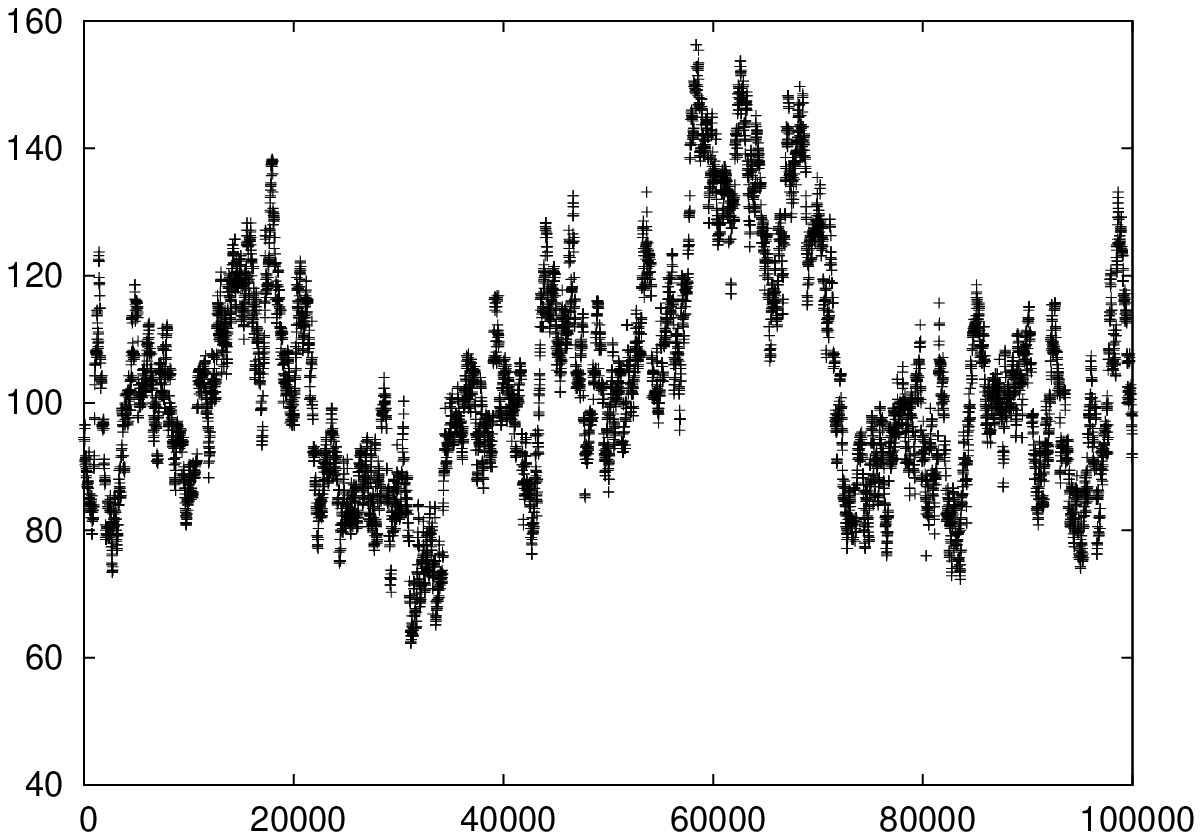}
\hfill
\includegraphics[width=.46\textwidth]{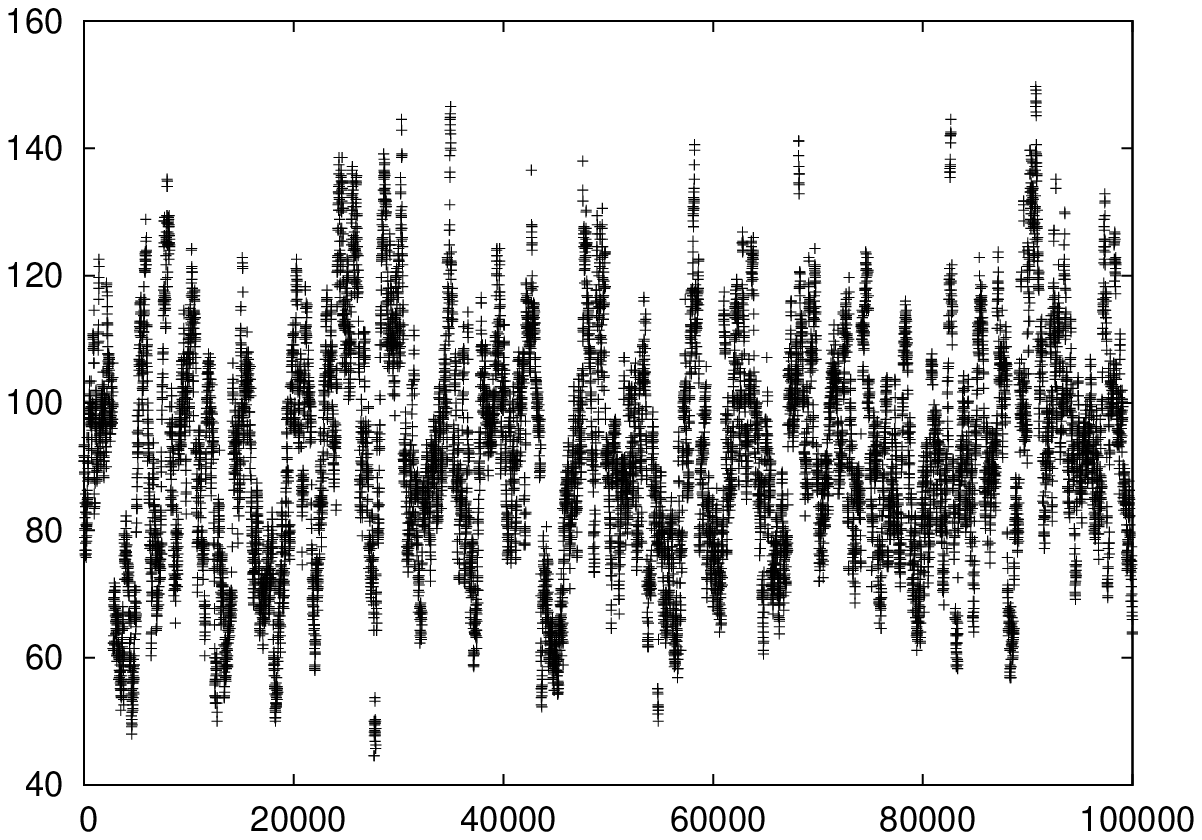}
\caption{\label{en}The first $100000$ numerical measurements of the interaction energy (a measurement takes place every $100$ MC steps) with the worm setup (left) and the diagonal setup (right). Strong autocorrelations are visible in the worm setup.}
\end{figure*}

\begin{figure*}
\includegraphics[width=.47\textwidth]{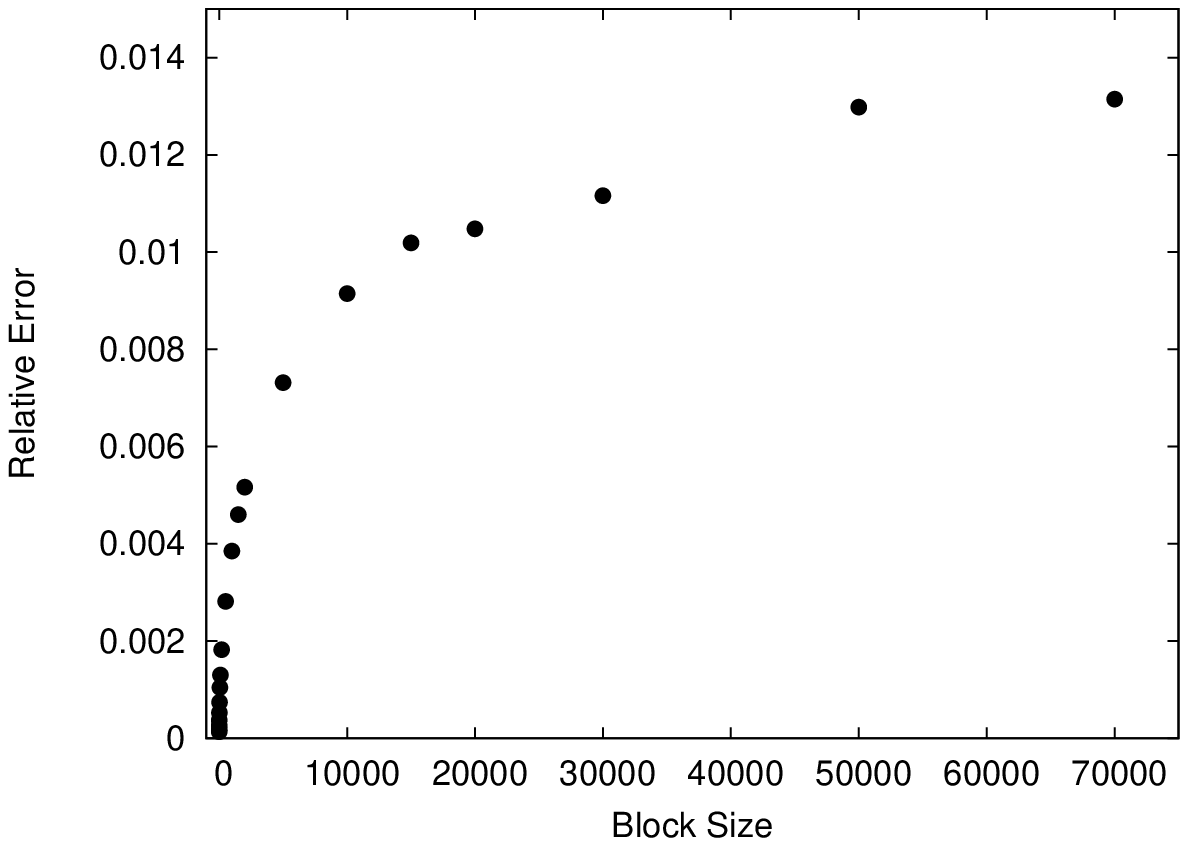}
\hfill
\includegraphics[width=.47\textwidth]{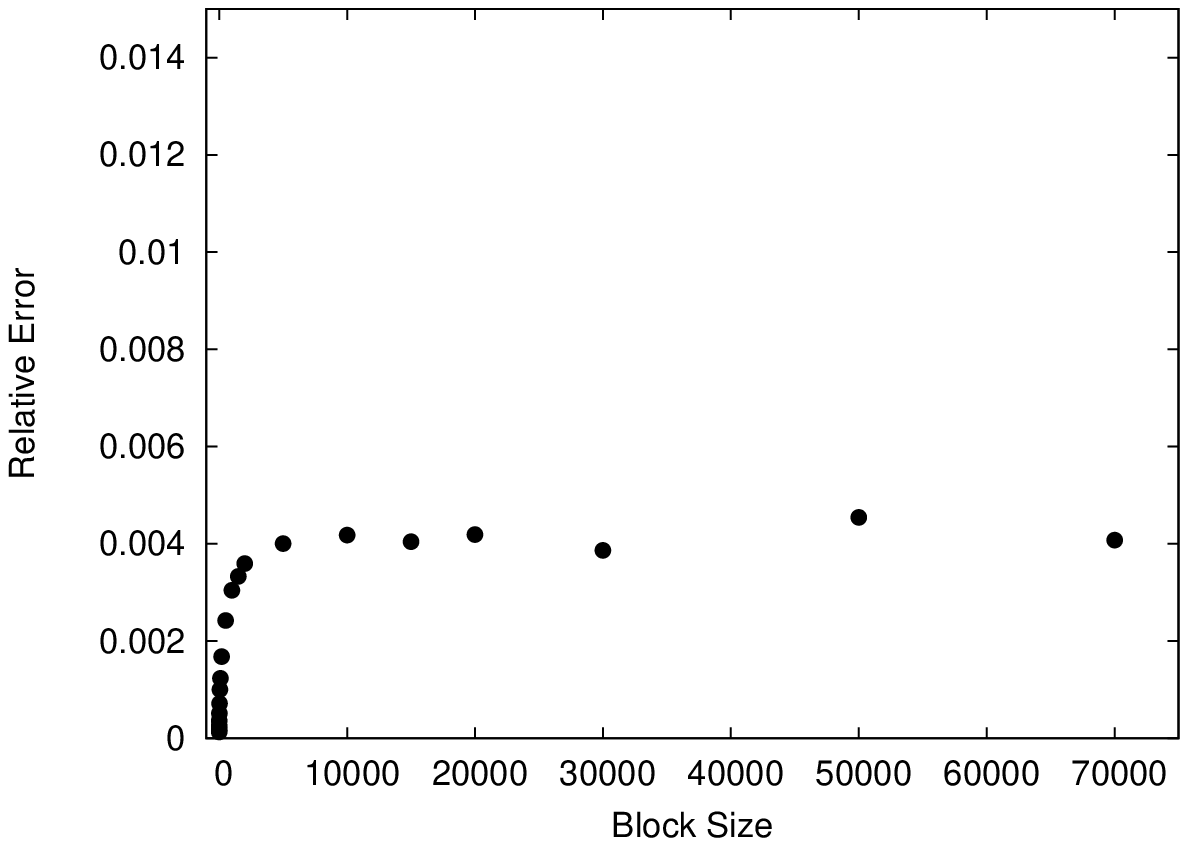}
\caption{\label{err}The blocking error analysis of the interaction energy with the worm setup (left) and the diagonal setup (right). The blocked error is much higher in the worm setup and continues increasing even for large block sizes.}
\end{figure*}

A detailed description of the individual updates can be found in the appendix of \cite{burovski}. The idea behind the worm algorithm is that at low densities the major contribution comes from multi-ladder diagrams, these are configurations where the vertices are arranged into several vertex chains. Proposing updates that favour the creation of such vertex chains will lead to higher acceptance ratios and increase the efficiency of the simulation. At low densities the acceptance ratios of the worm updates are an order of magnitude higher than the acceptance ratios of the simple diagonal updates, in which the vertices are inserted or removed at random. We found however, that the worm type addition and removal updates from the original setup suffer from strong autocorrelations, so that even after many successful updates the configuration does not change significantly \cite{goulko}. To illustrate this we compare the measurements of the interaction energy (which is proportional to the diagram order) in the worm setup and the diagonal setup in Fig.~\ref{en}. Both simulations used the same parameters and a comparable number of MC steps. Figure \ref{err} shows the blocking analysis of the relative error for the same quantity. Blocking is a widely used technique to estimate the error of an autocorrelated measurement. The single data points are arranged consecutively into blocks of equal size, and each block is replaced by the average of the measurements it contains. Then the error is calculated for the resulting blocked system in the usual way. If no autocorrelations are present, the error will be independent of the block size. In the presence of autocorrelations $N$ consecutive data points fluctuate less than $N$ independent measurements. Hence the error will increase with block size, until the block size reaches the autocorrelation length of the system.

\begin{figure}
\includegraphics[width=\columnwidth]{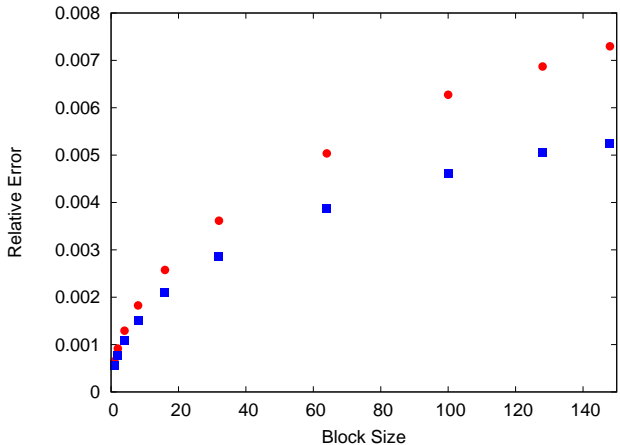}
\caption{\label{comparebalancedmodified}(Color online) The blocking error analysis of the interaction energy. Red circles correspond to the pure diagonal setup and blue squares to a combination of diagonal and modified worm updates with equal probabilities for each kind of update.}
\end{figure}

Because of the large errors due to autocorrelations the worm setup is effectively less efficient than the standard diagonal setup. For this reason in the present study we employ the conventional diagonal updates, together with the modified worm addition and removal updates, as proposed in \cite{goulko}. This setup combines the advantages of the diagonal setup (weak autocorrelations) with the ones of the worm setup (high acceptance ratios). Below is a summary of all updates used in our simulation. For the modified updates we also give the values of the acceptance ratios $\mathcal{R}$ (the other acceptance ratios can be found in \cite{burovski}). The corresponding formulae for the imbalanced case will be given in Sec.~\ref{imbalancedsection}.
\\\\
\textbf{Updates only concerning the worm vertices:}
\begin{itemize}
\item \textbf{Worm creation/annihilation:} insert/remove the pair $P(\mathbf{x},\tau),\ P^\dagger(\mathbf{x}',\tau')$ into/from the configuration. In our setup the distributions for $P$ and $P^\dagger$ are independent: both are distributed uniformly over the lattice, so that $W(S_p\rightarrow \tilde{S}_p)=(\beta L^3)^{-2}$, where $\tilde{S}_p$ stands for the configuration $S_p$ with the additional 2-point vertices. The authors of \cite{burovski} describe a setup in which the vertex $P$ is selected randomly, and the vertex $P^\dagger$ is then chosen in a spacetime hypercube of given extent around $P$. To avoid autocorrelations that can be associated with this scheme we employ the independent setup. The acceptance ratio is then
\begin{equation}
\mathcal{R}=\left|\frac{\det\mathbf{A}(\tilde{S}_p)}{\det\mathbf{A}(S_p)}\right|^2(\beta L^3)^2\zeta.
\end{equation}
\item \textbf{Worm shift:} shift the $P^\dagger(\mathbf{x}',\tau')$ vertex to other coordinates. This update is equivalent to the worm shift update in \cite{burovski} and involves a shift to a nearest neighbour on the lattice and a time shift in some interval around the old coordinates.
\end{itemize}
\textbf{Updates of the regular 4-point vertices:} adding/removing a 4-point vertex (changes the diagram order).
\begin{itemize}
\item \textbf{Diagonal version:} add or remove a random vertex. This is the most basic setup for changing the diagram order, however at low densities the acceptance ratios are very low.
\item \textbf{Modified worm-type updates:}
\begin{itemize}
\item Choose a random 4-point vertex from the configuration (which will act as a worm for this step).
\item Addition: add another 4-point vertex on the same lattice site and in some time interval of length $\Delta\tau$ around the worm.
\item Removal: remove the nearest neighbour of the worm vertex (implies that addition can only be accepted if the new vertex is the nearest neighbour of the worm).
\end{itemize}
The probability density for the addition update is then $W(S_p\rightarrow S_{p+1})=1/(p\Delta\tau)$, where $1/p$ comes from selecting the worm and $1/\Delta\tau$ from choosing the new time coordinate. Analogously for the removal update $W(S_p\leftarrow S_{p+1})=1/(p+1)$ and the acceptance ratio becomes
\begin{equation}
\mathcal{R}=\left|\frac{\det\mathbf{A}(S_{p+1})}{\det\mathbf{A}(S_p)}\right|^2\frac{(-U)p\Delta\tau}{p+1}.
\end{equation}
\end{itemize}
The modified worm setup still prolongs existing vertex chains like the original worm setup, but autocorrelations are significantly reduced since the worm changes with every update. This new type of updates can only be employed in addition to the regular diagonal addition and removal updates. It works regardless if the pair of 2-point vertices is present or not in the configuration (the original worm addition/removal updates can only take place when the 2-point vertices are present). The acceptance rates for this update are comparable with those for the regular worm updates. To demonstrate the increase in efficiency we compare the diagonal and the modified worm setup at low density, when the acceptance rates of the diagonal updates are particularly poor. We again consider the blocking analysis of the relative error for the interaction energy. As Fig.~\ref{comparebalancedmodified} clearly shows, the autocorrelation length does not increase in presence of the modified worm updates. The blocked error in this case is significantly lower due to the increased acceptance rate.

The structure of the updates requires the calculation of a matrix determinant of a large matrix (with rank $p$ up to about $6000$) in each MC step. Since only a few elements of the matrix change in the course of one update (at most one row and one column) a recalculation of the whole determinant from scratch is not necessary. In our implementation we make use of the fast matrix update formulae \cite{rubtsov} which decrease the number of required operations to order $p^2$ instead of order $p^3$.

\subsection{Imbalanced case\label{imbalancedsection}}
\begin{figure}
\includegraphics[width=\columnwidth]{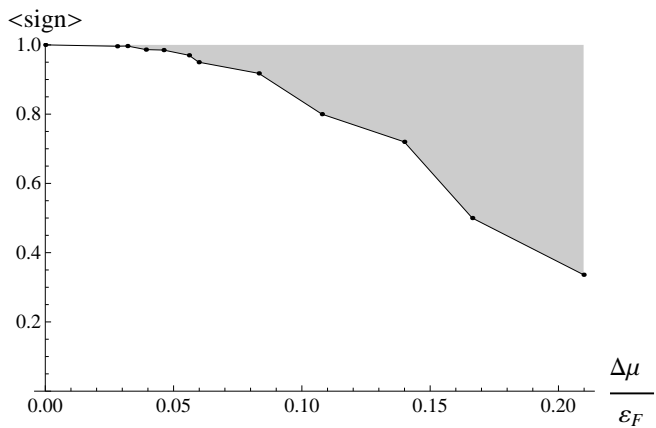}
\caption{\label{signplot}Schematic plot of the average sign near the critical point as a function of imbalance. The shaded area covers the range of values the sign can take at different values of lattice size and chemical potential. The lower boundary of this area is the ``worst-case'' curve of the sign, corresponding to lowest densities and largest lattice sizes used.}
\end{figure}

The original DDMC algorithm relies strongly on the assumption of equal densities of the two fermion species. This assumption allows us to write the partition function (\ref{partitionfunction}) as a sum of positive terms only, and consequently to use it as a probability distribution for Monte Carlo sampling. To study the imbalanced case $\mu_\uparrow\neq\mu_\downarrow$ a generalisation of the algorithm is necessary. Due to the sign problem the function $\rho_W(S_p)=\frac{1}{Z_W}(-U)^p\det\mathbf{A}^\uparrow(S_p)\det\mathbf{A}^\downarrow(S_p)$ is no longer positive for all configurations $S_p$ and can thus not be used as a probability distribution. To deal with this problem we will make use of the ``sign quenched method'', which is based on the ``phase quenched method'' known from lattice QCD \cite{phasequenched}. The idea is to write the function $\rho_W$ as a product of its modulus and its sign,
\begin{equation}
\rho_W(S_p)=\frac{1}{Z_W}(-U)^p|\det\mathbf{A}^\uparrow(S_p)\det\mathbf{A}^\downarrow(S_p)|\sgn(S_p),
\end{equation}
and to use the positive function
\begin{equation}
\rho'_W(S_p)\equiv\frac{1}{Z'_W}(-U)^p|\det\mathbf{A}^\uparrow(S_p)\det\mathbf{A}^\downarrow(S_p)|
\end{equation}
as the new probability distribution. The factor $Z'_W$ ensures normalisation. This reweighting implies another change in the Monte Carlo estimator for a generic thermodynamic observable $\langle\hat{X}\rangle$. When sampling according to the sign quenched probability distribution $\rho'_W(S_p)$, relation (\ref{MCestbalext}) becomes
\begin{equation}
\langle\hat{X}\rangle=\frac{\left\langle\mathcal{Q}^{(X,Z'_W)}\right\rangle_{\rho'_W}}{\left\langle\mathcal{Q}^{(Z,Z'_W)}\right\rangle_{\rho'_W}}=\frac{\left\langle\mathcal{Q}^{(X,Z_W)}(S_p)\sgn(S_p)\right\rangle_{\rho'_W}}{\left\langle\mathcal{Q}^{(Z,Z_W)}(S_p)\sgn(S_p)\right\rangle_{\rho'_W}}.
\end{equation}
The Monte Carlo estimators $\mathcal{Q}$ remain unchanged, apart from a multiplication with $\pm1$ depending on the relative sign of the two matrix determinants $\det\mathbf{A}^\uparrow(S_p)$ and $\det\mathbf{A}^\downarrow(S_p)$. This representation of a thermal average in terms of the new probability distribution is mathematically equivalent to the usual thermal average. However, numerical errors can become very large if the expectation value of the sign in the denominator is close to zero, as it happens for the expectation value of the phase in QCD. For the unitary Fermi gas the sign remains very close to unity for small imbalances, as shown in Fig.~\ref{signplot}, so that sign quenching is applicable for imbalances up to approximately $0.2\varepsilon_F$.
The restricting factor that keeps us from reaching large imbalances is not the sign, but rather the fact that even large values of $\Delta\mu$ do not necessarily lead to large differences in the filling factors of the two components and hence the physical value $\Delta\mu/\varepsilon_F$ still remains small. This method works best close to the balanced limit and can provide a useful tool to examine the trend of the critical temperature for small deviations from it.

The worm updates and acceptance ratios now generalise straightforwardly to the imbalanced case. In all formulae we merely need to replace the terms $|\det\mathbf{A}|^2$ by $|\det\mathbf{A}^\uparrow\det\mathbf{A}^\downarrow|$. A slight drawback is that we now need to keep in memory two large matrices instead of one and update each of these matrices separately. Also the relative error of the sign adds to the relative error of each observable.

\section{Results\label{resultssection}}
\subsection{Balanced results}
\begin{figure}
\includegraphics[width=\columnwidth]{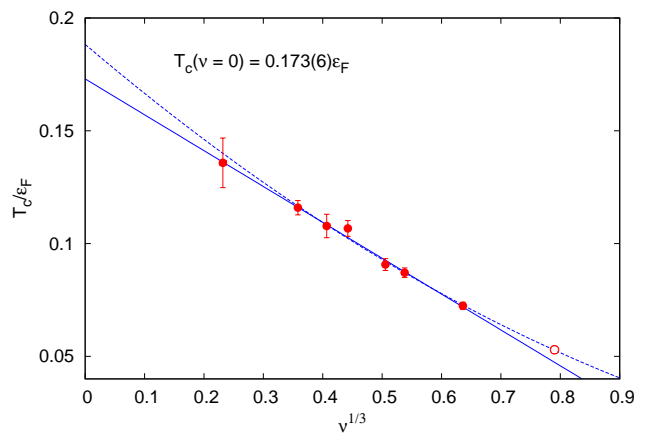}
\caption{\label{balancedtc}(Color online) The critical temperature versus filling factor for different values of the chemical potential. The continuum limit corresponds to $\nu\rightarrow0$. The linear extrapolation (solid line) of the seven data points at lowest filling factors (filled circles) yields a continuum value of $T_c/\varepsilon_F=0.173(6)$. The dashed line corresponds to a quadratic fit through all data points.}
\end{figure}

\begin{figure*}
\includegraphics[width=.47\textwidth]{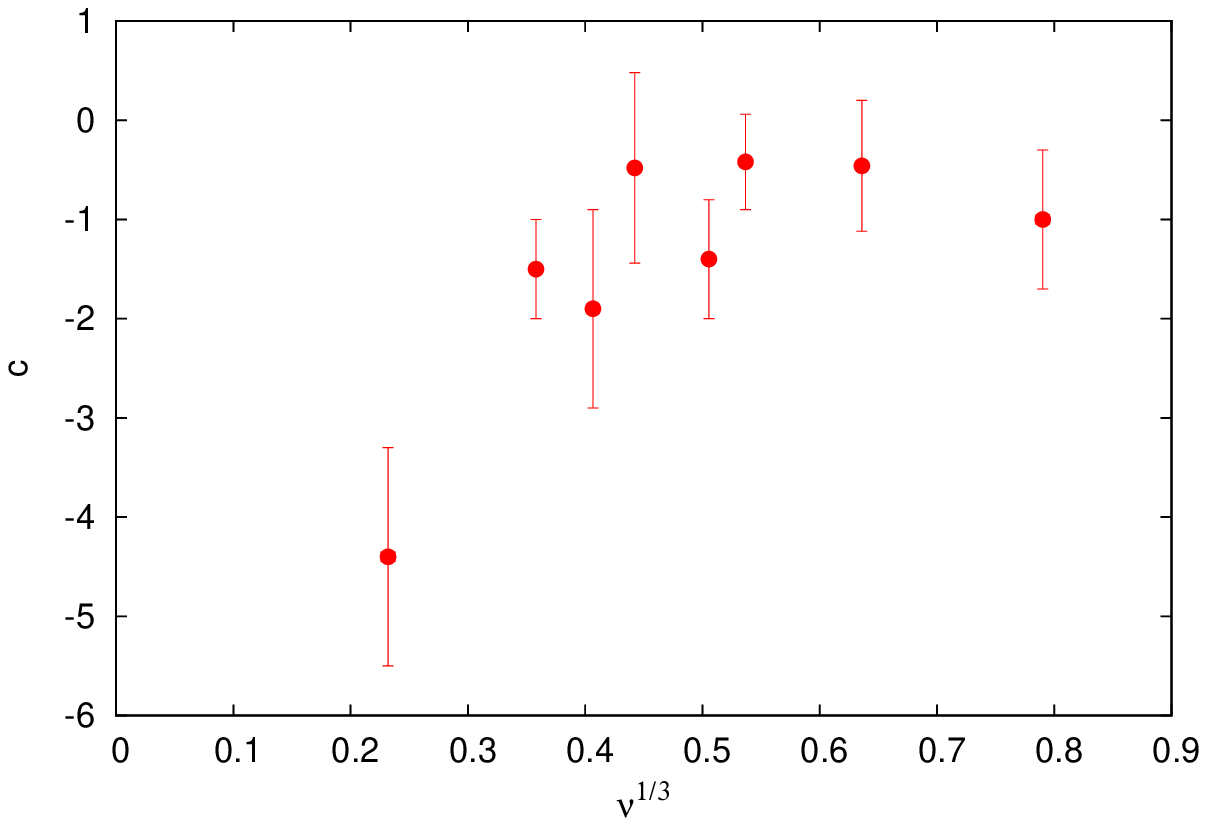}
\hfill
\includegraphics[width=.47\textwidth]{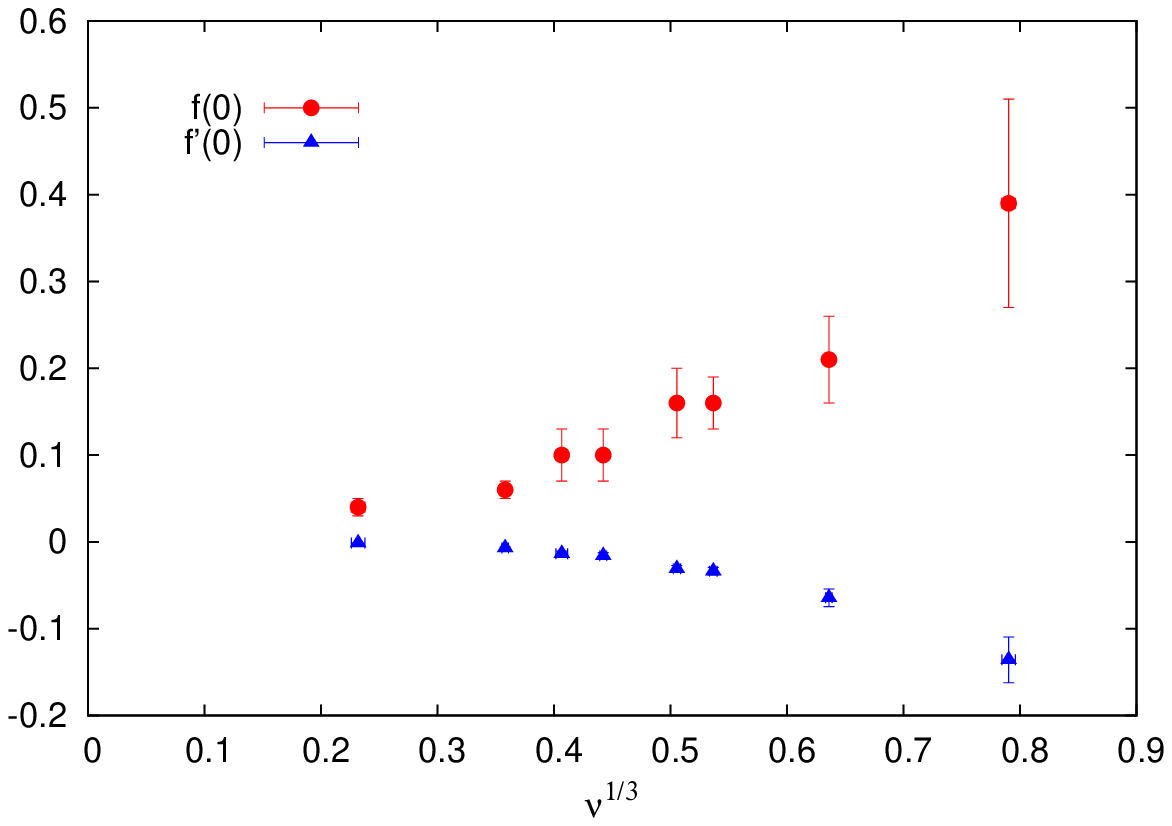}
\caption{\label{latticecoeffs}(Color online) The non-universal constant $c$ (left) and the first two coefficients in the expansion of the universal scaling function $f(x)$ (right) in lattice units versus filling factor, see Eq.~(\ref{Rscalingexp}) with $f(0)=f_0$ and $f'(0)=f_1T_c$.}
\end{figure*}

Before we include the imbalanced data we first present our analysis of the data at zero imbalance, for comparison with previous results from \cite{burovski}. We performed simulations at eight different values of the chemical potential, corresponding to eight different filling factors. The lattice sizes varied between $4^3$ for the highest filling factor and $26^3$ for the lowest, so that the volume range in physical units was approximately constant. As discussed in \cite{burovski}, for sufficiently small $\nu$ the critical temperature scales linearly with $\nu^{1/3}$. This behaviour is seen for $\nu^{1/3}\lessapprox0.75$. Our results and the continuum extrapolation are shown in Fig.~\ref{balancedtc}. A line was fitted through the seven points with $\nu^{1/3}<0.75$, resulting in $T_c/\varepsilon_F=0.173(6)-0.16(1)\nu^{1/3}$. The goodness of fit is $\chi^2/$d.o.f $=0.39$. For comparison we also fit a quadratic through all eight data points, resulting in a continuum value of $T_c/\varepsilon_F=0.188(15)$, which is in excellent agreement with the linear extrapolation. This confirms that sub-leading corrections proportional to $\nu^{2/3}$ can indeed be neglected for sufficiently small $\nu$. In Fig.~\ref{latticecoeffs} we show the results for the fit parameters $c$, $f_0$ and $f_1T_c$, according to Eq.~(\ref{Rscalingexp}). These parameters are smooth functions of the filling factor. This data shows that the non-universal constant $c$ does indeed take values of order unity and thus cannot be neglected.

\begin{figure*}
\includegraphics[width=.49\textwidth]{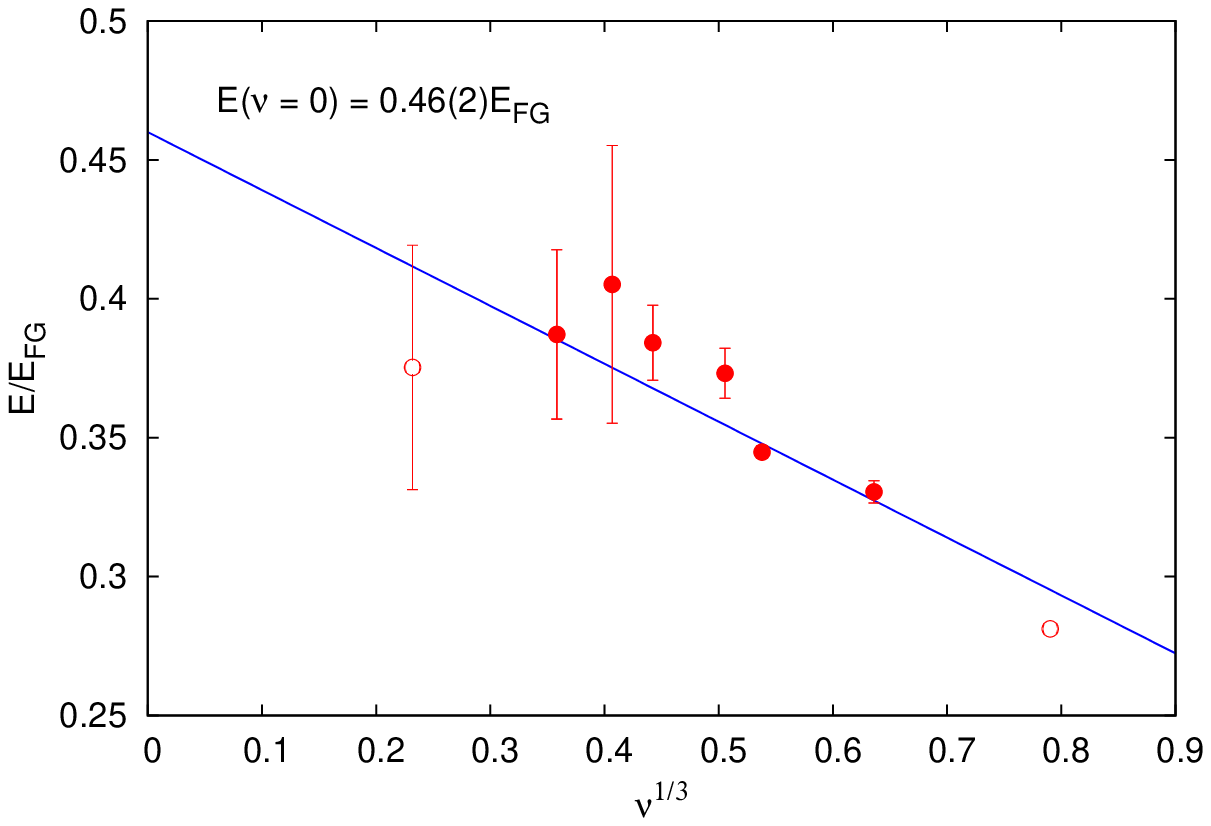}
\hfill
\includegraphics[width=.49\textwidth]{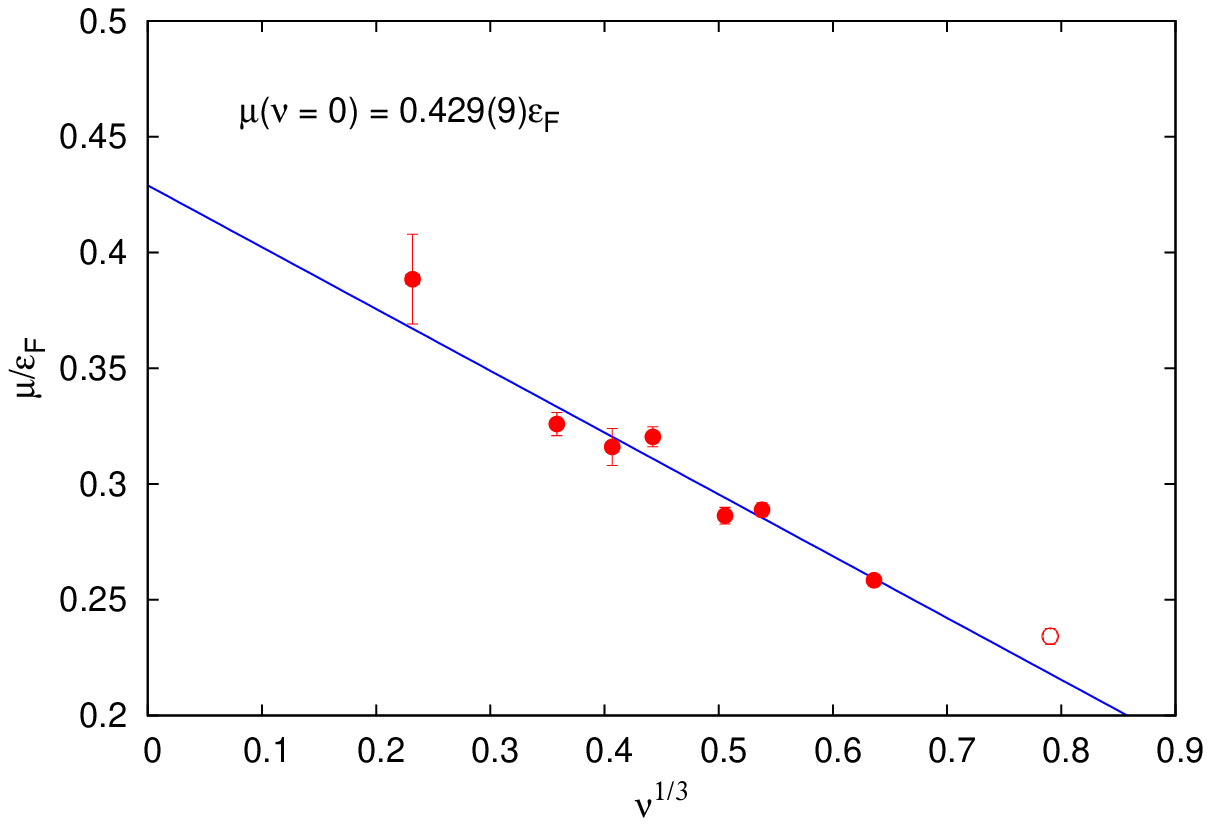}
\caption{\label{balancedotherobs}(Color online) The energy per particle (left) and the chemical potential (right) versus filling factor at the critical point. The linear fits were performed for data at filling factors $\nu^{1/3}<0.75$. Filled circles indicate data included in the fit and empty circles stand for data excluded from the fit, see text for discussion.}
\end{figure*}

Our final result for the critical temperature in physical units is thus $T_c/\varepsilon_F=0.173(6)$. This value is significantly higher than the previous result from \cite{burovski}, where $T_c/\varepsilon_F=0.152(7)$. We will also make a comparison with other results available from the literature. In \cite{burovskinew} the result of \cite{burovski} was found to be in agreement with a continuous space-time DDMC method. The authors of \cite{bulgac1} found an upper bound of $T_c/\varepsilon_F\lessapprox0.15(1)$. They used an auxiliary field Monte Carlo approach and extracted the critical temperature from the finite-size scaling of the condensate fraction, using the same procedure as in \cite{burovski}. The difference between our results might be attributed to the approximation made through this fitting method. A less recent result by the same group \cite{bulgacold1, bulgacold2} is $T_c/\varepsilon_F=0.23(2)$. Through extrapolating Monte Carlo results of low-density neutron matter, the authors of \cite{abeTc} found a value of $T_c/\varepsilon_F=0.189(12)$ at unitarity. Their value agrees with our result within errors. There are also results obtained with the Restricted Path Integral Monte Carlo method \cite{rpimc}, $T_c/\varepsilon_F\approx0.245$, and an upper bound of $T_c/\varepsilon_F<0.14$ obtained with a hybrid Monte Carlo method \cite{lee}. Results obtained with an $\epsilon$-expansion are also available \cite{epsexp}. For comparison, the critical temperature in the BEC limit is $T_{\textnormal{BEC}}=0.218\varepsilon_F$.

The authors of \cite{wernercastin} conjecture that the leading order change of the critical temperature is linear in $k_Fr_e$, where $k_F=\sqrt{\varepsilon_F}$ is the Fermi wavevector and $r_e$ is the effective range of the potential ($r_e=-0.3056$ in units of lattice spacing for the Fermi-Hubbard model \cite{revcastin}), with a model independent coefficient. Our result for the linear slope is $\Delta T_c/\varepsilon_F=-0.16(1)$, which is larger in magnitude than the value from \cite{burovski}.

A similar continuum extrapolation can be performed for other thermodynamic observables, like the energy per particle and the chemical potential. The corresponding data together with the fits is presented in Fig.~\ref{balancedotherobs}. The total energy is composed of the kinetic energy $E_{\textnormal{kin}}$ and the interaction energy $E_{\textnormal{int}}=\langle H_1\rangle$. An explicit expression for the former can be obtained from the position space picture,
\begin{eqnarray}
E_{\textnormal{kin}}&=&-\left\langle\sum_{\mathbf{x},\sigma}c^\dagger_{\mathbf{x}\sigma}\nabla^2c_{\mathbf{x}\sigma}\right\rangle\\
&=&-\left\langle \sum_{\mathbf{x},\sigma}c^\dagger_{\mathbf{x}\sigma}\sum_{j=1}^3(c_{(\mathbf{x}+\hat{\mathbf{j}})\sigma}+c_{(\mathbf{x}-\hat{\mathbf{j}})\sigma}-2c_{\mathbf{x}\sigma})\right\rangle\\
&=&2\langle L^3(6c^\dagger_{\mathbf{x}}c_{\mathbf{x}}-6c^\dagger_{\mathbf{x}}c_{\mathbf{x}+\hat{\mathbf{j}}})\rangle,\ \textnormal{for any}\ \hat{\mathbf{j}},\;\sigma.
\end{eqnarray}
The factor $L^3$ in the last line comes from summation over all lattice sites and the other numerical factors are due to summation over $j$ and $\sigma$. Since $2\langle c^\dagger_{\mathbf{x}}c_{\mathbf{x}}\rangle$ corresponds to the filling factor, the kinetic energy per particle can be written as
\begin{equation}
E_{\textnormal{kin}}/L^3\nu=6(1-2\langle c^\dagger_{\mathbf{x}}c_{\mathbf{x}+\hat{\mathbf{j}}}\rangle/\nu).
\label{energyestimator}
\end{equation}
The Monte Carlo estimator for the interaction energy is given by $\mathcal{Q}^{(H_1)}(S_p)=-\beta^{-1}p$, as shown in \cite{burovski}.

The results shown in Fig.~\ref{balancedotherobs} are obtained at $T_c$, but the temperature dependence of the chemical potential and the energy per particle was found to be very weak. Also these quantities showed almost no dependence on lattice size $L$. For the energy per particle we obtain the continuum value $E/N\varepsilon_F=0.276(14)$. In units of the ground state energy of the free gas, $E_{FG}=(3/5)N\varepsilon_F$, our result is $E/E_{FG}=0.46(2)$. Since the expression for the kinetic energy (\ref{energyestimator}) involves a difference of two quantities of comparable size, large fluctuations can occur, especially at low filling factors. For this reason measurements of the energy per particle at lowest filling factor could only be performed on lattices with size $L\leq14$, which is smaller than the lattice sizes used for the measurement of the critical temperature. We include this point in the plot in Fig.~\ref{balancedotherobs} (left), but exclude it from the linear fit. The goodness of fit is $\chi^2/$d.o.f. $=2.1$. Our result shows excellent agreement with the value $E/E_{FG}=0.45(1)$ at $T_c$ quoted in \cite{bulgac1}. The value quoted in \cite{burovski} is $E/N\varepsilon_F=0.31(1)$, which in units of the free ground state energy roughly corresponds to $E/E_{FG}=0.52(2)$.

For the chemical potential at $T_c$ we obtain the continuum value $\mu/\varepsilon_F=0.429(9)$ with $\chi^2/$d.o.f. $=2.8$. Our result differs from the value $\mu/\varepsilon_F=0.493(14)$ quoted in \cite{burovski}, but is consistent with the value $\mu/\varepsilon_F=0.43(1)$ quoted in \cite{bulgac1}.

Since the chemical potential and the energy are expected to stay almost constant at temperatures below $T_c$ we also make a comparison to values from the literature obtained at zero temperature. In the zero temperature limit the quantities $\mu/\varepsilon_F$ and $E/E_{FG}$ are equal and Monte Carlo estimates range between approximately $0.40(1)$ and $0.44(1)$ \cite{zeromu1, zeromu2, zeromu3, zeromu4}. Our value for the chemical potential falls within this range, the value for the total energy is slightly higher, which is consistent with the fact that the energy must increase at finite temperature. These numerical estimates are consistent with experiment \cite{zeromuexp1, zeromuexp2, zeromuexp3}.

Finally, we make a comparison with recent experimental studies of the homogeneous unitary Fermi gas. A direct measurement of the critical temperature and the chemical potential of the uniform gas has been presented in \cite{salomon}. Their experimental value $T_c/\varepsilon_F=0.157(15)$ agrees well with our result. However, the value of the chemical potential at the critical point $\mu/\varepsilon_F=0.49(2)$ differs from our value. Another experimental determination of the critical temperature and thermodynamic functions, including the energy and the chemical potential, is described in \cite{horikoshi}. Their values $T_c/\varepsilon_F=0.17(1)$ and $\mu/\varepsilon_F=0.43(1)$ at $T_c$ show excellent agreement with our results. Their result for the energy per particle $E/N\varepsilon_F=0.34(2)$ at $T_c$ is higher than our value. In another experimental work \cite{MITexp} an estimate for the critical temperature at zero imbalance is extrapolated from data at higher values of imbalance.

\subsection{Imbalanced Results}
\begin{figure}
\includegraphics[width=\columnwidth]{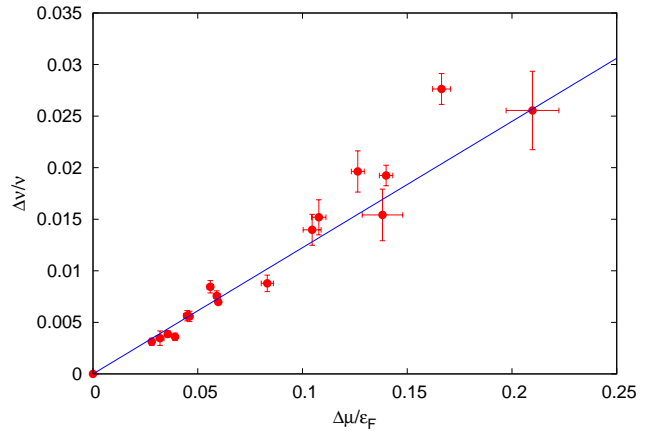}
\caption{\label{munu}(Color online) Relation between the chemical potential difference and the relative density difference at $T_c$.}
\end{figure}

Now we will present our results for the imbalanced case $\mu_\downarrow\neq\mu_\uparrow$. Data was taken at $25$ points, of which $23$ lie within the regime of linear scaling, $\nu^{1/3}<0.75$. Out of these $23$ points $7$ are at zero imbalance, as discussed in the previous section. The two most common ways of quantifying imbalance are either through the chemical potential difference $\Delta\mu/\varepsilon_F=|\mu_\uparrow-\mu_\downarrow|/\varepsilon_F$, or through the relative density difference $\Delta\nu/\nu=|\nu_\uparrow-\nu_\downarrow|/(\nu_\uparrow+\nu_\downarrow)$. For the values of imbalance considered in our study these two quantities are proportional to each other, with $\Delta\nu/\nu=0.122(2)\Delta\mu/\varepsilon_F$, as illustrated in Fig.~\ref{munu}. The relative density difference shows no dependence on lattice size (the $L$-dependencies of $\nu$ and $\Delta\nu$ cancel each other out), but considerable dependence on the temperature. Also since $\Delta\nu$ is a small quantity, numerical fluctuations can become significant. Since the chemical potential difference is less prone to numerical errors, we will use it from now on to quantify imbalance.

The critical temperature $T_c/\varepsilon_F$ is now a function of filling factor $\nu$ and imbalance $h=\Delta\mu/\varepsilon_F$ (in the following we assume that all quantities are in physical units and do not always write the $\varepsilon_F$ factors explicitly). We are ultimately interested in the continuum limit corresponding to $\nu=0$ and want to perform the corresponding extrapolation. To achieve this all numerical data is fitted to a three dimensional surface, where the following assumptions are made for the form of the fitted function:
\begin{itemize}
\item At fixed imbalance the critical temperature is a linear function of $\nu^{1/3}$, with slope $\alpha(h)$: $T_c(\nu,h=\textnormal{const})=T_c(h)+\alpha(h)\nu^{1/3}$. This is a generalisation of the relation valid in the balanced case.
\item $T_c(h)$ and $\alpha(h)$ viewed as functions of the imbalance $h$ are analytic and can thus be Taylor expanded.
\item Due to symmetry in $h$ all odd powers in the Taylor expansions of $T_c(h)$ and $\alpha(h)$ have to vanish.
\item $T_c(h)$ must be a non-increasing function of $h$.
\end{itemize}
Hence the fitted function takes the form
\begin{equation}
T_c(\nu,h)=T_c(h)+\alpha(h)\nu^{1/3}.
\end{equation}
If we expand $T_c(h)$ and $\alpha(h)$ to leading order in $h$ the fitted function becomes
\begin{equation}
T_c(\nu,h)=T_0+T_2h^2+(\alpha_0+\alpha_2h^2)\nu^{1/3}.
\label{imbquad}
\end{equation}
This requires a linear fit of four parameters. The best fit yields $T_0=0.171(5)$, $\alpha_0=-0.154(9)$, $T_2=0.4\pm0.9$ and $\alpha_2=-0.7\pm1.9$ with $\chi^2/$d.o.f.$=0.43$. Note that the $T_2$ value corresponding to the minimal $\chi^2$ is positive, which is forbidden by physical arguments. The $\chi^2$ function is very flat along the $T_2$ direction, so that forcing $T_2=0$ results in $\chi^2/$d.o.f.$=0.44$. From the error on $T_2$ we derive the lower bound $T_2>-0.5$. The best fit values for $T_0$ and $\alpha_0$ are in excellent agreement with the ones obtained from the fit of the balanced data only.

The error on the best fit value for $\alpha_2$ is very large and the fit is consistent with $\alpha_2=0$. Hence we also perform a fit to the function
\begin{equation}
T_c(\nu,h)=T_0+T_2h^2+\alpha_0\nu^{1/3},
\label{imbquadconst}
\end{equation}
where $T_c(h)$ has again been expanded to quadratic order and the function $\alpha(h)$ has been replaced by a constant $\alpha_0$. The best fit is
\begin{equation}
T_c(\nu,h)=0.171(5)+0.07(11)h^2-0.155(8)\nu^{1/3},
\end{equation}
with $\chi^2/$d.o.f.$=0.41$. This $\chi^2$-value is even lower than for the previous fit, which means that the data justifies dropping the $\alpha_2$ term. The best fit result is still consistent with $T_2=0$ and leads to a much tighter lower bound $T_2>-0.04$. The other parameters $T_0$ and $\alpha_0$ agree with the results from the previous fit and the fit of the balanced data.
\begin{figure}
\includegraphics[width=\columnwidth]{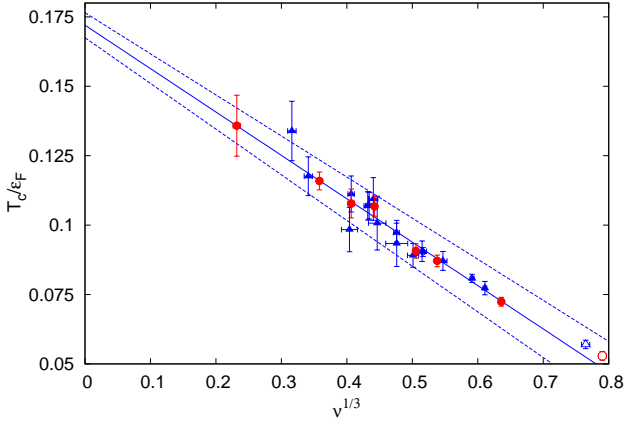}
\caption{\label{projT}(Color online) Projection of the data onto the $(\nu^{1/3}$-$T_c)$ plane. Red circles denote the balanced data and blue triangles data at non-zero imbalance. The line corresponds to the constant fit (\ref{imbconstnumbers}). Dashed lines denote the error margins.}
\end{figure}
\begin{figure}
\includegraphics[width=\columnwidth]{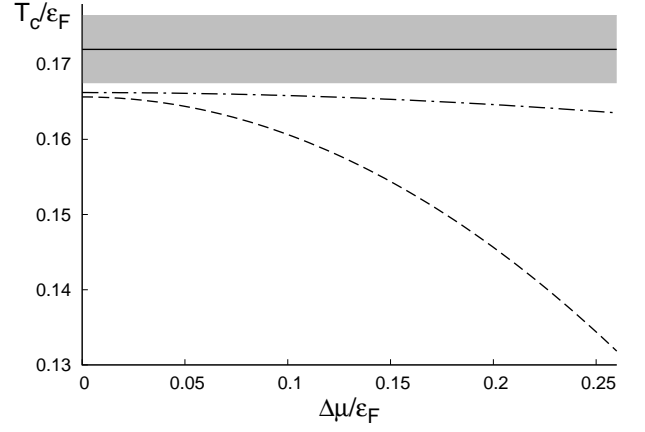}
\caption{\label{tcbounds}The continuum limit of the critical temperature as a function of imbalance. The solid line is the value obtained from the constant fit (\ref{imbconst}), the shaded area corresponds to one standard deviation. The dashed line is the lower bound obtained from fit (\ref{imbquad}) and the dot-dashed line is the tighter lower bound obtained from fit (\ref{imbquadconst}).}
\end{figure}
\begin{figure}
\includegraphics[width=\columnwidth]{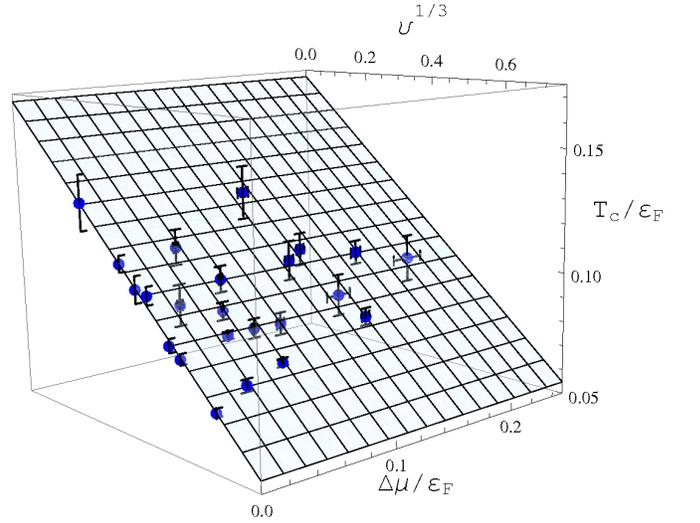}
\caption{\label{3Dtemp}(Color online) Three dimensional plot of the critical temperature versus filling factor and imbalance. The surface corresponds to the constant fit (\ref{imbconstnumbers}).}
\end{figure}

Since our results indicate that $T_c$ remains almost unchanged in response to a weak imbalance, we also perform a fit to constant $T_c(h)$ and $\alpha(h)$,
\begin{equation}
T_c(\nu,h)=T_0+\alpha_0\nu^{1/3}.
\label{imbconst}
\end{equation}
This is the same function as the one used in the balanced case and corresponds to a straight line fitted through the projection of all data points onto the $(\nu^{1/3}$-$T_c)$ plane, see Fig.~\ref{projT}. The best fit is
\begin{equation}
T_c(\nu,h)=0.1720(45)-0.156(8)\nu^{1/3},
\label{imbconstnumbers}
\end{equation}
with $\chi^2/$d.o.f.$=0.41$. Again the result agrees with the previous fits.

\begin{table*}
\caption{Comparison of fit parameters obtained by different fit methods.\label{fitcomptable}}
\begin{ruledtabular}
\begin{tabular}{lccccccccc}
& $T_0$ & $\delta T_0$ & $T_2$ & $T_2$ lower bound & $\alpha_0$ & $\delta\alpha_0$ & $\alpha_2$ & $\delta\alpha_2$ &$\chi^2$/d.o.f. \\
\hline
balanced data & 0.173 & 0.006 & & & -0.16 & 0.01 & & & 0.39 \\
fit to eq. (\ref{imbquad}) & 0.171 & 0.005 & 0.4 & -0.5 & -0.154 & 0.009 & -0.7 & 1.9 & 0.43 \\
fit to eq. (\ref{imbquadconst}) & 0.171 & 0.005 & 0.07 & -0.04 & -0.155 & 0.008 & & & 0.41 \\
fit to eq. (\ref{imbconst}) & 0.172 & 0.0045 & & & -0.156 & 0.008 & & & 0.41
\end{tabular}
\end{ruledtabular}
\end{table*}
\begin{figure}
\includegraphics[width=\columnwidth]{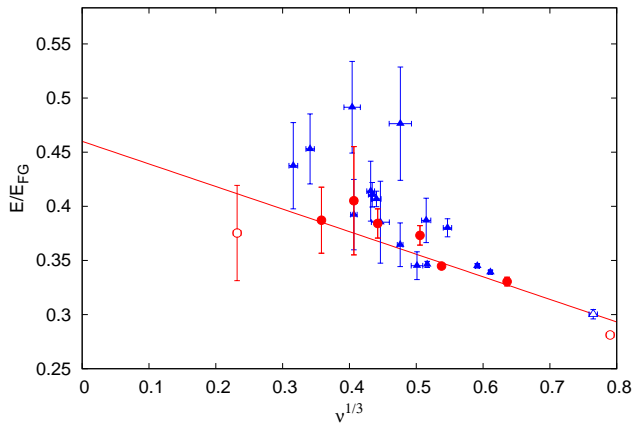}
\caption{\label{projenergy}(Color online) Projection of the data for the energy per particle onto the $(\nu^{1/3}$-$E)$ plane. Red circles denote the balanced data and blue triangles data at non-zero imbalance. For comparison the fit at zero imbalance is shown. The points at non-zero imbalance tend to lie above the balanced fit line.}
\end{figure}
\begin{figure}
\includegraphics[width=\columnwidth]{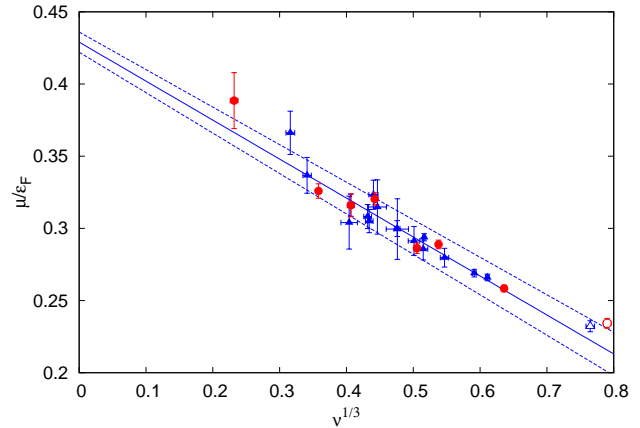}
\caption{\label{projchempot}(Color online) Projection of the data for the average chemical potential onto the $(\nu^{1/3}$-$\mu)$ plane. Red circles denote the balanced data and blue triangles data at non-zero imbalance. The solid line corresponds to the constant fit and the dashed lines indicate the error margins.}
\end{figure}
We also performed fits using the jackknife method and several robust fits. All results were consistent with the minimal $\chi^2$ fits. Table \ref{fitcomptable} provides an overview of the results obtained with the different fit methods. The values for the parameters $T_0$ and $\alpha_0$ were obtained with high accuracy and are all in excellent agreement with each other, independently of the form of the fit function. Depending on the model assumptions two lower bounds could be derived for the leading order deviation of the critical temperature from its balanced value. Figure \ref{tcbounds} shows these two bounds compared with the value in the balanced case. A three dimensional plot of the data together with a constant surface fit is presented in Fig.~\ref{3Dtemp}. Experimental determinations of $T_c$ as a function of $h$ are an area of active research, and much larger values of $h$ can be reached \cite{MITexp}.

A similar analysis was performed for the energy per particle and the average chemical potential $\mu/\varepsilon_F=|\mu_\uparrow+\mu_\downarrow|/2\varepsilon_F$. Since with increasing imbalance interactions become suppressed, we expect the absolute value of the interaction energy to decrease. This in turn means an increase of the total energy, since the interaction energy is negative. As we did for the critical temperature we fit the energy in units of $E_{FG}$ to the function
\begin{equation}
E(\nu,h)=E_0+E_2h^2+(\alpha^{(E)}_0+\alpha^{(E)}_2h^2)\nu^{1/3}
\end{equation}
and obtain the best fit parameters $E_0=0.440(15)$, $\alpha^{(E)}_0=-0.17(3)$, $E_2=3.4\pm2.2$ and $\alpha^{(E)}_2=-3.1\pm4.5$, with $\chi^2/$d.o.f.$=2.8$. These results are consistent with the balanced fit. The leading coefficient $E_2=3.4\pm2.2$ is no longer consistent with zero. We also perform a fit to the function
\begin{equation}
E(\nu,h)=E_0+E_2h^2+\alpha^{(E)}_0\nu^{1/3}
\end{equation}
and obtain the best fit result
\begin{equation}
E(\nu,h)=0.444(13)+1.9(3)h^2-0.18(2)\nu^{1/3},
\end{equation}
which agrees with the previous result. The $\chi^2/$d.o.f.$=2.7$. Figure \ref{projenergy} shows the numerical data.

The average chemical potential is not expected to depend on the imbalance. Hence we fit our data to the function
\begin{equation}
\mu(\nu,h)=\mu_0+\alpha^{(\mu)}_0\nu^{1/3}
\end{equation}
and obtain
\begin{equation}
\mu(\nu,h)=0.429(7)-0.27(1)\nu^{1/3},
\end{equation}
with $\chi^2/$d.o.f.$=1.1$. This is in very good agreement with our balanced result. A plot of the data and the fit is in Fig.~\ref{projchempot}.

\section{Summary and Conclusion}
In this paper we presented a Monte Carlo calculation of the critical temperature and other thermodynamic observables of the unitary Fermi gas with equal and unequal chemical potentials for the two spin components. For our study we developed a modified version of the worm algorithm, which is less susceptible to autocorrelations than the previously available methods. This algorithm can be generalised to the imbalanced case using the sign quenched method. We first calculated the value of the critical temperature using only data at zero imbalance and found $T_c=0.173(6)\varepsilon_F$, which is significantly higher than the result previously obtained with the worm algorithm \cite{burovski}. One possible explanation is the difference in finite size analysis methods as described in Sec.~\ref{ordersection}. Our results for the energy and the chemical potential are $E=0.46(2)E_{FG}$ and $\mu=0.429(9)\varepsilon_F$.

In the imbalanced case we extracted the dependence of the critical temperature on the imbalance $h=\Delta\mu/\varepsilon_F$ close to the balanced limit. Our analysis is consistent with $T_c/\varepsilon_F=\textnormal{const.}$ for the range of imbalances considered ($h\lessapprox0.2$). The value at $h=0$ extracted from a quadratic fit of the balanced and imbalanced data was found to be $T_c(h=0)=0.171(5)\varepsilon_F$, which agrees with the value obtained from the balanced data only. We further derived a lower bound on the leading order term in the expansion of the critical temperature $T_c(h)-T_c(0)>-0.5\varepsilon_F$. With the additional assumption that the linear dependence of $T_c/\varepsilon_F$ on $\nu^{1/3}$ remains unchanged in the presence of a small imbalance a tighter lower bound of $T_c(h)-T_c(0)>-0.04\varepsilon_F$ could be obtained. We also analysed the behaviour of the energy and the chemical potential in the presence of an imbalance.

\begin{acknowledgments}
We are grateful to Evgeni Burovski and Boris Svistunov for the interesting and productive discussions. This work has made use of the resources provided by the Cambridge High Performance Computing Facility. OG is supported by the German Academic Exchange Service (DAAD), the Engineering and Physical Sciences Research Council (EPSRC) and the Cambridge European Trust.
\end{acknowledgments}

\bibliography{imb_apsstyle}

\end{document}